\documentclass{svproc}
\usepackage{url}
\usepackage{graphicx}
\usepackage{multicol}
\usepackage{footmisc}
\usepackage{caption}
\usepackage{amsmath}
\usepackage{bm}  %
\usepackage{ulem}
\usepackage{mwe}
\usepackage{comment}
\usepackage{verbatim}
\usepackage[titletoc,title]{appendix}
\usepackage{comment}
\usepackage{wrapfig}
\usepackage{lipsum}
\usepackage{floatrow}
\usepackage{epstopdf}

\begin{document}
\mainmatter              
\title{Dark Matter Bound States from Three-Body Recombination}
%
%
\author{Eric Braaten\inst{1} \and Daekyoung Kang\inst{2} \and Ranjan Laha\inst{3}}
%
%
\tocauthor{Eric Braaten, and Daekyoung Kang, and Ranjan Laha}
\institute{Department of Physics, The Ohio State University, Columbus, OH 43201, USA\\
\email{braaten.1@osu.edu},
\and
Key Laboratory of Nuclear Physics and Ion-beam Application (MOE) and 
Institute of Modern Physics, 
Fudan University, Shanghai, China 200433\\
\email{dkang@fudan.edu.cn},
\and
PRISMA Cluster of Excellence and
             Mainz Institute for Theoretical Physics, 
             Johannes Gutenberg-Universit\"{a}t Mainz, 55099 Mainz, Germany\\
\email{ranjalah@uni-mainz.de}
}

\maketitle              

\begin{abstract}
The small-scale structure problems of the universe can be solved by self-interacting dark matter
that becomes strongly interacting at low energies.
A particularly predictive model  is resonant short-range self-interactions, with
a dark-matter mass of about 19~GeV and a large S-wave scattering length of about 17~fm.
Such a model makes definite predictions for the few-body physics of weakly bound clusters of the dark-matter particles.
We calculate the production of two-body bound clusters by three-body recombination in the early universe
under the assumption that the dark matter particles are identical bosons, which is the most favorable case
for forming larger clusters.  The fraction of dark matter in the form of two-body bound clusters 
can increase by as much as 4 orders of magnitude
when the dark-matter temperature falls below the binding energy, but its  present value remains less than $10^{-6}$.
\end{abstract}

 \keywords{Dark matter, bound states, large scattering length, early universe.}

\section{Introduction}

The model with collisionless cold dark matter and  a cosmological constant
provides an excellent description of the large-scale structure of the universe,
but it has encountered problems at smaller scales associated with galaxies and clusters of galaxies.
The problems involve the dark-matter distribution  in the cores of galaxies
and the properties of  satellite galaxies.
They can all be solved by self-interacting dark matter that is strongly interacting at low energies \cite{Tulin:2017ara}.

\begin{figure}[t]
\includegraphics*[width=0.8\linewidth]{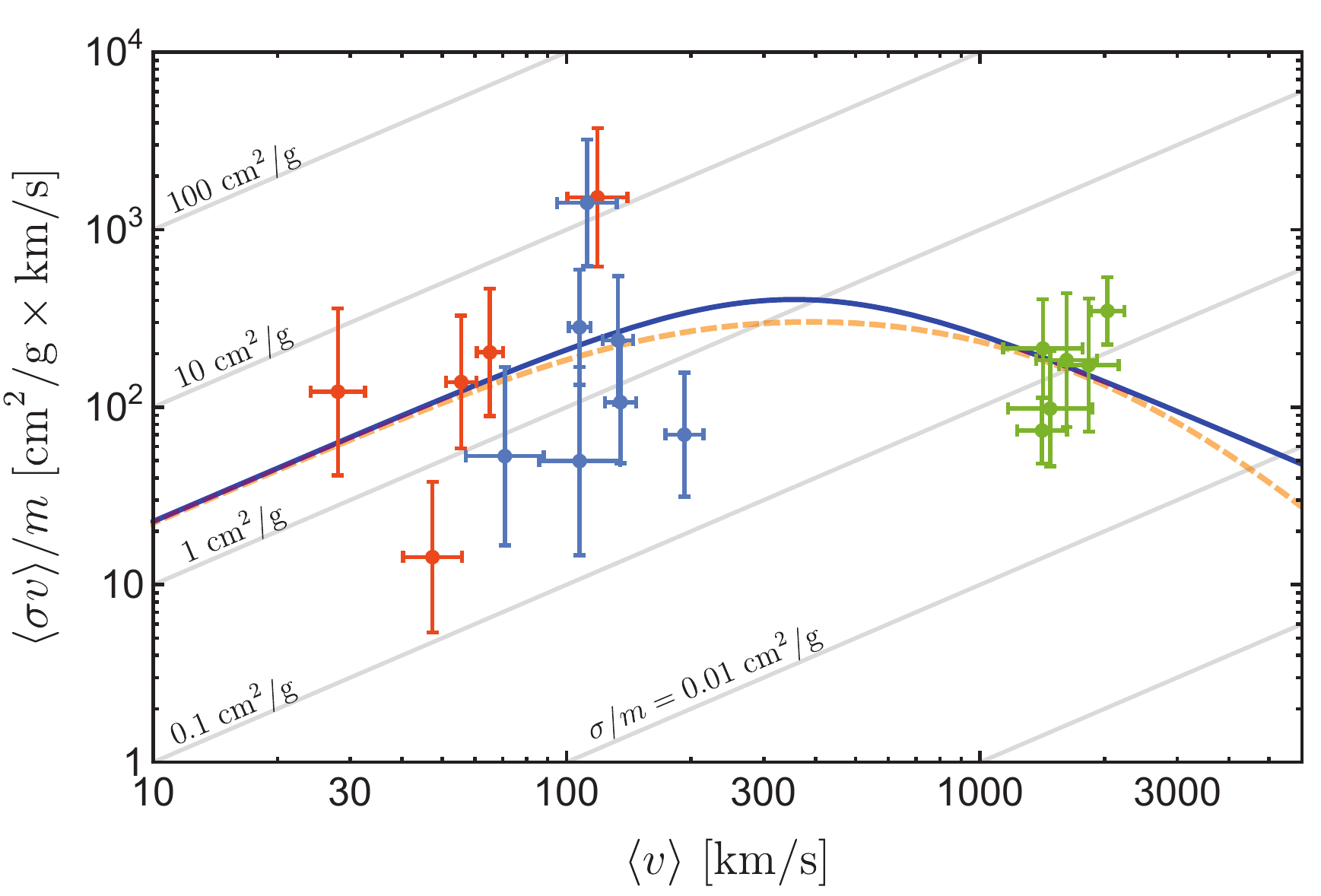} 
\caption{
Self-interaction reaction rate $\langle v\, \sigma_\mathrm{elastic}\rangle$ for dark matter particles 
as a function of  their  mean velocity $\langle v\rangle$ (adapted from \cite{Braaten:2018xuw}).
The data points from Ref.~\cite{Kaplinghat:2015aga} are for dwarf galaxies (red), low-surface-brightness galaxies (blue), 
and galaxy clusters (green) \cite{Kaplinghat:2015aga}.
The curves are the best fits to the model of Ref.~\cite{Kaplinghat:2015aga}
(dashed) and to Eq.~\eqref{eq:vsigma} (solid).
The diagonal lines are energy-independent cross sections. }
\label{sigmav}
\end{figure}

In Ref.~\cite{Kaplinghat:2015aga},
Kaplinghat, Tulin, and Yu deduced self-interaction reaction rates $\langle v \sigma_\mathrm{elastic}\rangle$
and mean  velocities $\langle v \rangle$ of dark matter particles 
for a number of dwarf galaxies, low-surface-brightness galaxies, and clusters of galaxies.
Their results are shown in Fig.~\ref{sigmav}.  They fit their results with
a simple  self-interacting dark matter model with 3 parameters:
the dark matter mass $m_\chi$,  a dark mediator mass   $\mu$, and a  Yukawa coupling $\alpha'$.
Their best fit for the masses with fixed Yukawa coupling $\alpha'  = 1/137$
was $m_\chi= 15$~GeV and $\mu = 17$~MeV \cite{Kaplinghat:2015aga}.

In Ref.~\cite{Braaten:2018xuw}, we showed that the results in Fig.~\ref{sigmav}
can be fit equally well by a simpler self-interacting dark matter model with 2 parameters.
The model has resonant short-range interactions 
with an S-wave resonance close to the scattering threshold \cite{Braaten:2013tza}.
The parameters are the dark matter mass $m_\chi$ and the scattering length $a$.  
This model has been applied previously to the direct detection of dark matter \cite{Laha:2013gva,Laha:2015yoa}.
The self-interaction reaction rate as a function of the velocity $v$ is
\begin{equation}
v\, \sigma_\mathrm{elastic}(v) =\frac{8 \pi a^2v}{1 + (a m_\chi/2)^2 v^2} .
\label{eq:vsigma}
\end{equation}
The best fit to the results in  Fig.~\ref{sigmav} is $m_\chi = 19$~GeV and $a= \pm 17$~fm \cite{Braaten:2018xuw}.

The dark matter could all be in the form of individual dark matter particles $d$,
but some (or all) of it could be bound into few-body clusters $d_N$,
which we call {\it dark nuclei}.
There are two basic formation mechanisms for larger dark nuclei.
If there is a light mediator $\gamma_d$ for dark matter self-interactions,
larger dark nuclei can be formed by radiative reactions:  $d + d_{N-1}  \to  d_N + \gamma_d$.
If there is no light mediator, as in our resonant short-range interaction model,
larger dark nuclei must be formed instead by  rearrangement reactions,
      such as 3-body recombination:   $d + d + d_{N-1}  \to  d_N + d$.
The formation of dark deuterons $d_2$ is a bottleneck for formation of larger dark nuclei $d_N$.

\section{Few-body Physics}
\label{two-body}

The low-energy  two-body physics of particles with resonant short-range interactions is very simple.
It is completely determined by the large scattering length $a$.
The cross section for the elastic scattering reaction  $d + d  \to  d + d$ is given in Eq.~\eqref{eq:vsigma}.
If $a$ is negative, there are no 2-body bound states.
If $a$ is positive, there is a single 2-body bound state $d_2$ that we call the dark deuteron.
Its binding energy is $E_2 = 1/m_\chi a^2$.

If the particles are identical bosons, the 3-body physics is much more intricate \cite{Braaten:2018xuw}.
It is determined not only by the large scattering length $a$, but also by a 3-body parameter.
There is a sequence of 3-body bound states called  Efimov states.
In the limit $a \to \pm\infty$, there are infinitely many Efimov states
with an accumulation point at the 3-boson threshold
and with the binding energy of each successive Efimov state smaller by a factor of $22.7^2 = 515$.
Three-boson reaction rates also have remarkable behavior.
They depend  log-periodically on a 3-body parameter $a_+$ with discrete scaling factor 22.7.
If $a > 0$, a simple example is the rate for the 3-body recombination reaction  $d + d + d  \to  d_2 + d$
at 0 collision energy:
\begin{equation}
R(E=0) = \frac{399.8\, \sin^2[s_0 \log(a/a_+)]}{1- 0.00717\,\sin^2[s_0 \log(a/a_+)]}\,a^4/m_\chi,
\label{eq:RE=0}
\end{equation}
where $s_0 = 1.00624$.  The  3-body recombination rate at nonzero collision energy $E$
has been calculated in Ref.~\cite{SEGB:2002} and in Ref.~\cite{Braaten:2008kx}.

\begin{figure}[t]
\centering
\includegraphics[width=0.8\linewidth]{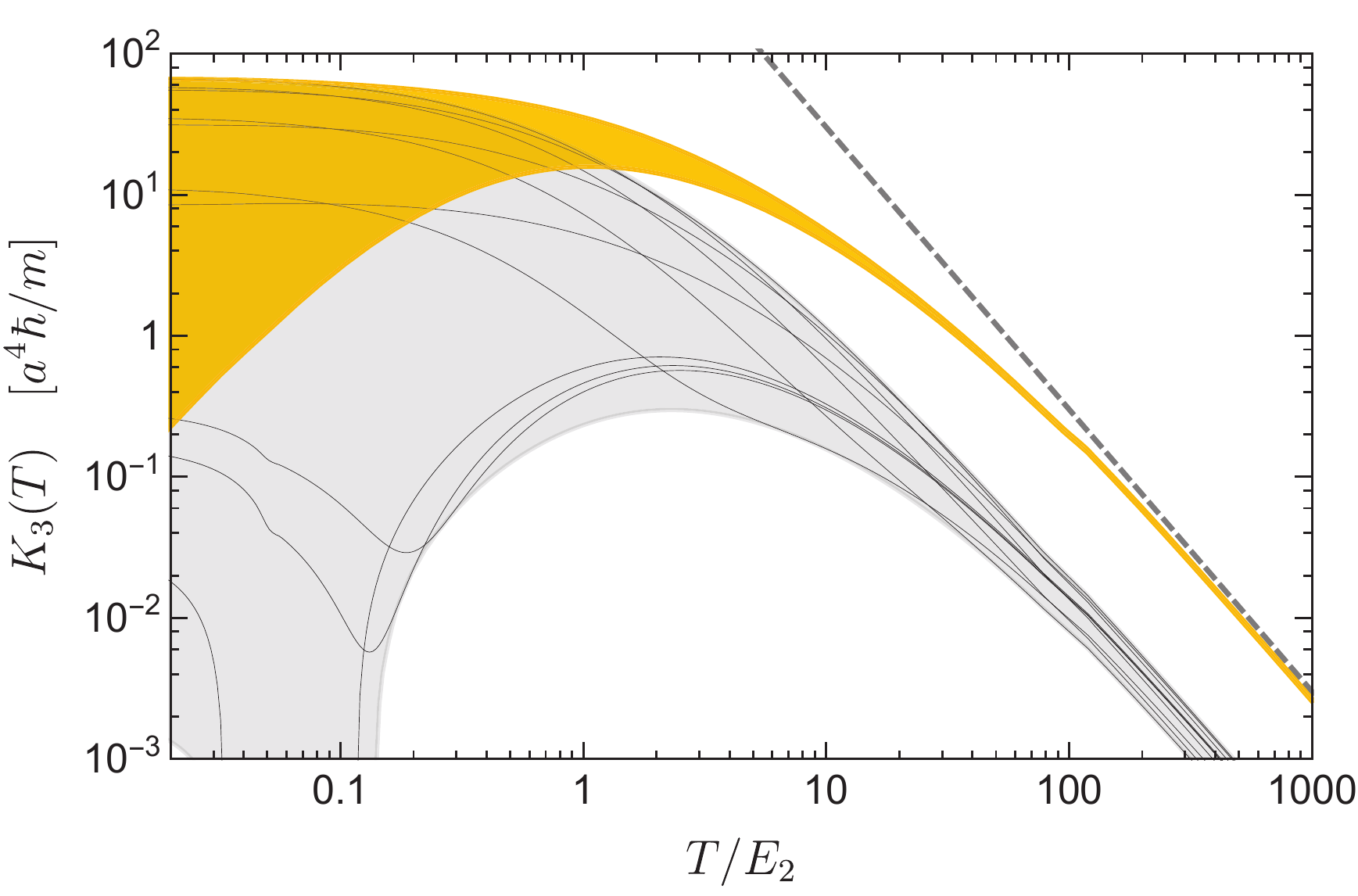}
\caption{
Rate coefficient $K_3(T)$ for three-body recombination as a function of the temperature $T$  
(adapted from \cite{Braaten:2018xuw}).
The upper band is the  envelope of $K_3(T)$ for all possible values of  the three-body parameter $a_+$.
The dashed line is the extrapolation from the scaling behavior at high temperature.
The lower  band is the envelope of the $J=0$ contribution to $K_3(T)$ for all possible  values of $a_+$. 
The curves inside the lower band are for 8 values of $a_+$.}
\label{fig:K3}
\end{figure}

We consider a gas consisting of dark matter particles $d$ with number density $n_1$
and dark deuterons $d_2$ with number density $n_2$ in thermal equilibrium at temperature $T$.
The rate of change in the number density of dark deuterons is 
\begin{equation}
\frac{d\ }{dt} n_2 = + K_3(T) \, n_1^3  - K_2(T) \, n_1 n_2,
\label{eq:dn2/dt:K3}
\end{equation}
where $K_3(T)$ and $K_2(T)$ are the rate constants for 3-body recombination   
and for the dark deuteron breakup reaction  $d + d_2  \to  d + d + d$.
These rate constants were calculated in Ref.~\cite{Braaten:2008kx}. 
The results for $K_3(T)$ are shown in Fig.~\ref{fig:K3}.
                                                       
\section{Dark Matter in the Early Universe}
\label{dark matter}

We calculate the formation of dark deuterons in the Hubble expansion of the early universe,
taking into account the 3-body recombination and dark deuteron breakup reactions.
 We calculate the number densities $n_1$ and $n_2$ 
 as functions of the redshift $z$, which is a convenient time variable.
The initial condition is that $n_2$ is negligible when dark matter decouples  at a redshift  
of about $z_\text{dc}  \approx m_\chi/20kT_\mathrm{cmb}$,
where $T_\mathrm{cmb}$ is the present temperature of the cosmic microwave background \cite{Steigman:2012nb}.
For $m_\chi=19$~GeV, this redshift is $z_\text{dc} \approx 10^{13}$.
The dark-matter temperature  as a function of $z$ is
\begin{equation}
T(z)  = T_\mathrm{cmb}\,\frac{ (1 + z)^2}{1+z_\text{dc}}\,.
\label{eq:Tdark-z}
\end{equation}
The total number density of dark matter  is determined by the present mass density $\rho_\mathrm{cdm}$
of cold dark matter:
\begin{equation}
n_1(z) +2 n_2(z)  = \frac{\rho_\mathrm{cdm}}{m_\chi} \, (1+z)^3 \,.
 \label{eq:ndarkz}
\end{equation}
It is convenient to express our results in terms of the dark deuteron mass fraction:
\begin{equation}
f_2(z)  =2 \, n_2(z)/[n_1(z) + 2n_2(z)] \,. 
 \label{eq:dfraction}
\end{equation}

\begin{figure}[t]
\centering
\includegraphics[width=0.8\linewidth]{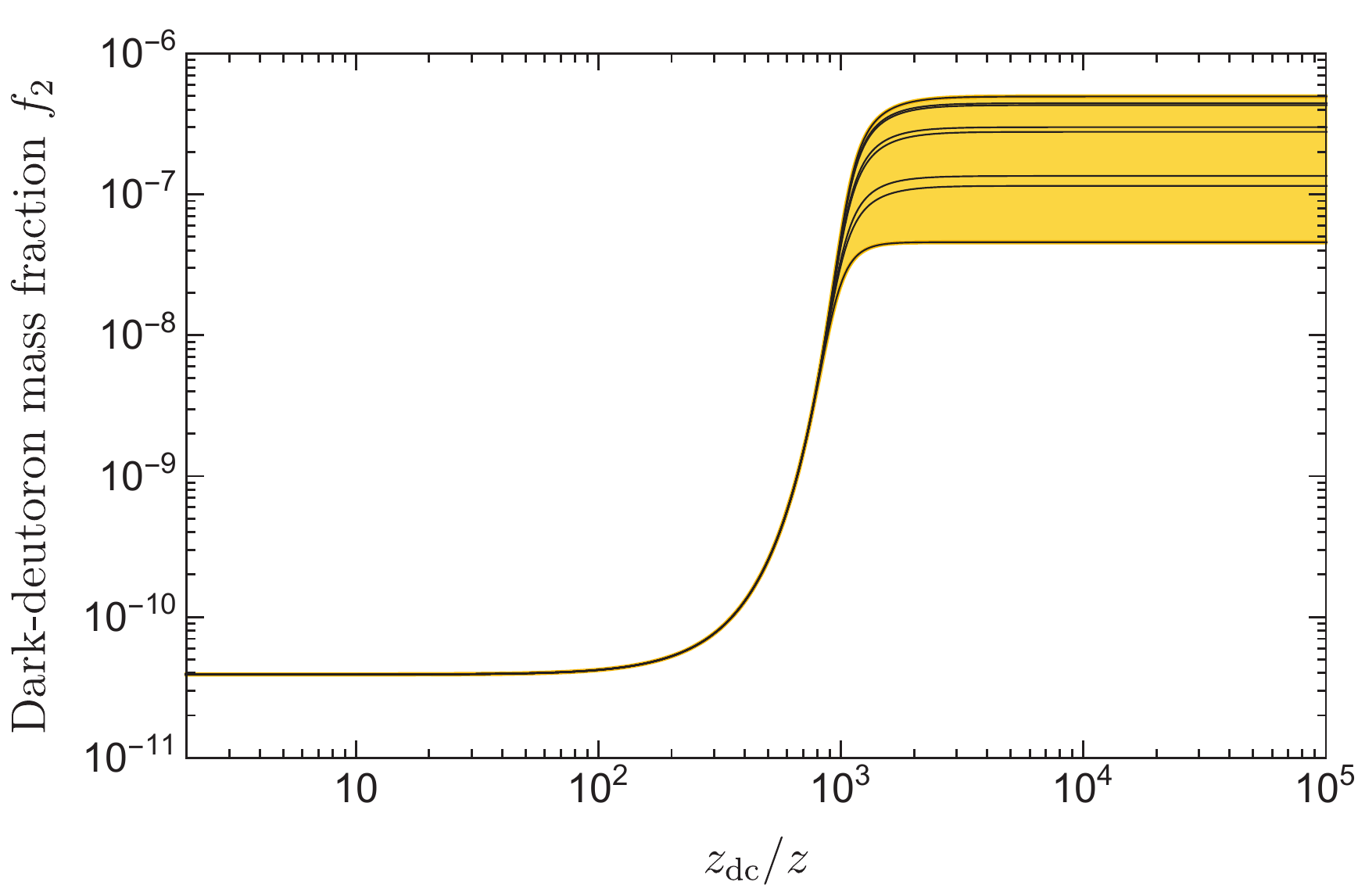}
\caption{Dark-deuteron mass fraction $f_2(z)$ as a function of 
$z_\text{dc}/z$ for $m_\chi = 19$~GeV and $a=17$~fm
(adapted from \cite{Braaten:2018xuw}).
The band is the envelope of all possible  values of $a_+$. 
The  curves are for 8 values of $a_+$.}
\label{fig:fdimer2}
\end{figure}

We assume the dark matter particles are identical bosons with mass $m_\chi = 19$~GeV, 
large scattering length $a=17$~fm, and unknown 3-body parameter $a_+$.
The dark deuteron mass fraction $f_2(z)$ is shown as a function of the redshift variable 
$z_\text{dc}/z$ in Fig.~\ref{fig:fdimer2}.
The fraction increases by 3 or 4 orders of magnitude around the redshift $10^{-3} z_\mathrm{dc}$
when $kT$ is equal to the binding energy $E_2 = 7$~keV  of the dark deuteron.
At smaller $1/z$, there is a  plateau in $f_2$ at about $4 \times 10^{-11}$
from equilibrium between recombination and breakup.
The final fraction $f_2(0)$ depends log-periodically on $a_+$,
ranging from $5\times10^{-8}$ to $5 \times 10^{-7}$.
It can be increased by ignoring the data from 
clusters of galaxies  in Fig.~\ref{sigmav}, which means  keeping $a=17$~fm but allowing $m_\chi$  to decrease.
The final fraction $f_2(0)$ can be increased to about $10^{-2}$ for $m_\chi=0.4$~GeV.
If $m_\chi$ is smaller, $f_2(0)$ is sensitive to the range of self-interactions.

In summary, the small-scale structure problems of the universe
can be solved by a self-interacting dark matter model with resonant S-wave interactions,
with parameters $m_\chi \approx 19$~GeV and  $a \approx \pm 17$~fm.
Dark nuclei $d_N$ must be produced by rearrangement reactions,
                                          such as 3-body recombination:  $d + d + d_{N-1}  \to  d_N + d$                                                                   
The most favorable case for producing dark nuclei larger than the dark deuteron is 
for the dark matter particles to be identical bosons. 
We found that a significant fraction of dark deuterons 
cannot be formed in the early universe by 3-body recombination.                                                                    
Since the formation of the dark deuteron $d_2$ is a bottleneck for the formation of larger dark nuclei $d_N$,
they cannot be formed either.

\section*{Acknowledgments}
This research was supported in part by U.S.~National Science Foundation, U.S.~Department of Energy,
Natural Science Foundation of China, German Research Foundation, and European Research Council.

%
%


\begin{thebibliography}{6}

\bibitem{Tulin:2017ara} 
Tulin, S.\ and Yu, H.B.:
  Phys.\ Rept.\  {\bf 730}, 1 (2018).
  
  
\bibitem{Kaplinghat:2015aga}
Kaplinghat, M., et al.:
  Phys.\ Rev.\ Lett.\  {\bf 116}, 041302 (2016).
  
\bibitem{Braaten:2018xuw}
Braaten, E., et al.:  
  JHEP {\bf 1811}, 084 (2018).
  
\bibitem{Braaten:2013tza} 
Braaten, E.\ and Hammer, H.-W.:
  Phys.\ Rev.\ D {\bf 88}, 063511 (2013).
  
\bibitem{Laha:2013gva} 
Laha, L.\ and Braaten, E.:
  Phys.\ Rev.\ D {\bf 89},   103510  (2014).
  
\bibitem{Laha:2015yoa} 
Laha, R.:
  Phys.\ Rev.\ D {\bf 92},  083509 (2015).
   
  \bibitem{SEGB:2002}
Suno, H., et al.:
Phys.\ Rev.\ A {\bf 65},  042725 (2002).
  
\bibitem{Braaten:2008kx} 
Braaten, E., et al.:
  Phys.\ Rev.\ A {\bf 78},   043605 (2008).

\bibitem{Steigman:2012nb}
Steigman, G., et al:
  Phys.\ Rev.\ D {\bf 86}, 023506 (2012).
  
\end{thebibliography}
\end{document}